\documentclass[prd,superscriptaddress,twocolumn,
floatfix,preprintnumbers,altaffilletter]{revtex4}
\usepackage{graphicx,url,amssymb,amsmath,longtable,rotating,color,units,
wasysym,subfigure,epsfig,multirow}
\usepackage{epstopdf}

\begin{document}
% DCC P1500093

\title{Detecting gravitational-wave transients at five sigma: \\a hierarchical approach}

\author{Eric~Thrane}
\email{eric.thrane@monash.edu}
\affiliation{School of Physics and Astronomy, Monash University, Clayton, Victoria 3800, Australia}

\author{Michael~Coughlin}
\affiliation{Department of Physics, Harvard University, Cambridge, MA 02138, USA}

\begin{abstract}
  As second-generation gravitational-wave detectors prepare to analyze data at unprecedented sensitivity, there is great interest in searches for unmodeled transients, commonly called bursts.
  Significant effort has yielded a variety of techniques to identify and characterize such transient signals, and many of these methods have been applied to produce astrophysical results using data from first-generation detectors.
  However, the computational cost of background estimation remains a challenging problem; it is difficult to claim a $5\sigma$ detection with reasonable computational resources without paying for efficiency with reduced sensitivity.
  We demonstrate a hierarchical approach to gravitational-wave transient detection, focusing on long-lived signals, which can be used to detect transients with significance in excess of $5\sigma$ using modest computational resources.
  In particular, we show how previously developed seedless clustering techniques can be applied to large datasets to identify high-significance candidates without having to trade sensitivity for speed.
\end{abstract}

\maketitle

{\em Introduction.}
With second-generation gravitational-wave (GW) detectors coming online later this year, the first direct detection of GWs may be near.
In order to establish the significance of a detection, it is common to report a false alarm probability (FAP), which quantifies the probability that a noise fluctuation could produce an event at least as loud as the observed candidate (as measured by some detection statistic).
%EHT: updated on 15 September, 2015 for one-sided spectrum
%In some subfields, e.g., particle physics,``$5\sigma$ significance,'' corresponding to $\text{FAP}\approx5.7\times10^{-7}$, is used as a detection threshold.
In some subfields, e.g., particle physics,``$5\sigma$ significance,'' corresponding to $\text{FAP}\approx2.9\times10^{-7}$, is used as a detection threshold.

In order to estimate the $\text{FAP}$ of a GW candidate, it is common to perform time-shifts in which the GW strain time series from one detector is shifted with respect to the series from a second detector by an amount greater than both the travel time between the detectors and the coherence time of the signals being targeted.
Time-shifting preserves non-Gaussian and non-stationary features that characterize the zero-lag (no time-shift) noise, while simultaneously eliminating true GW signals.
By performing $N$ time-shifts, it is possible to generate a distribution of the detection statistic, which can be used to estimate $\text{FAP}$ to a level of $\geq1/N$.
%EHT: updated on 15 September, 2015 for one-sided spectrum
%The $5\sigma$ threshold corresponds to $N\approx1.7\times10^6$.
The $5\sigma$ threshold corresponds to $N\approx3.5\times10^6$.
In many cases it is computationally impractical to carry out this many time-shifts, though, it has been accomplished in the ``detection'' of a LIGO blind injection with a matched filter search~\cite{s6lowmass}.
% 29 September, 2015
% removed one reference
Despite the pervasive use of time-shifts, there are limitations~\cite{0264-9381-27-1-015005}.

For many transient GW searches, the significant computing costs incurred by background estimation arise from a desire to use a coherent detection statistic.
Coherent algorithms utilize the complex-valued cross-power obtained by cross-correlating strain data from $\geq$$2$ detectors instead of or in addition to the incoherent auto-power observed in each detector separately; see, e.g.,~\cite{X-Pipeline,CoherentWaveBurst,stamp,s6vsr23_allsky}.
The extra phase information  helps differentiate signal from background, improving the sensitivity of the search.
However, the cost of background estimation for coherent searches is relatively large compared to a comparable incoherent search because the detection statistic must be recalculated for each time-shift, after the fresh application of a clustering algorithm.
Some algorithms use single-detector auto-power exclusively, which allows for much more rapid background estimation~\cite{box,s5vsr1_allsky}.

In this Letter, we describe a hierarchical approach to background estimation in the context of a search for long-lived, unmodeled GW transients using seedless clustering~\cite{stochtrack,stochsky}.
First, we identify ``events'' using a computationally intensive, but incoherent, single-detector statistic.
Second, we calculate a computationally fast, coherent detection statistic for each event identified with the single-detector statistic.
The second, coherent detection statistic is used to evaluate significance.
% 29 September, 2015
% shortened sentence
By splitting the calculation into an incoherent stage and a coherent stage, the computationally intensive calculations are carried out just once, allowing rapid background estimation without sacrificing the sensitivity gained by the use of coherence.

%EHT: updated on 15 September, 2015 for one-sided spectrum
%We demonstrate this technique by estimating the background---past the $5\sigma$ level---for two weeks of simulated Monte Carlo data and two weeks of Initial LIGO noise, recolored to resemble data from Advanced LIGO~\cite{aligo}.
We demonstrate this technique by estimating the background---past the $5\sigma$ level---for one week of simulated Monte Carlo data and one week of Initial LIGO noise, recolored to resemble data from Advanced LIGO~\cite{aligo}.
We calculate the sensitivity of this mock search for several toy model waveforms and find that it is not adversely affected by the incoherent stage.
% 29 September, 2015
% combine paragraphs
The remainder of this Letter is organized as follows.
%EHT: updated on 15 September, 2015 for one-sided spectrum
%We review the basics of transient identification with seedless clustering and describe the details of the new hierarchical detection statistic, we describe a mock data analysis carried out on two weeks of Monte Carlo data and two weeks of recolored Initial LIGO noise, and we present results demonstrating the ability to estimate background at the $5\sigma$ level.
We review the basics of transient identification with seedless clustering and describe the details of the new hierarchical detection statistic, we describe a mock data analysis carried out on one week of Monte Carlo data and one week of recolored Initial LIGO noise, and we present results demonstrating the ability to estimate background at the $5\sigma$ level.

{\em Method.}
In previous work, we have described seedless clustering~\cite{stochtrack,stochsky,verylong,stochtrack_cbc,stochtrack_ecbc}, a technique in which GW transients are identified by looking for clusters of excess coherence integrated along many different parametrized curves through frequency-time space.
In particular, cubic B\'ezier curves~\cite{bezier} provide a useful family of curves suitable for the detection of bursting sources lasting tens of seconds to weeks, which slowly evolve in frequency, but which are approximately narrowband on short time scales~\cite{stochtrack,stochsky,verylong}.
% 16 July, 2015: should piro:07 be moved into the accretion disk sentence?
Such long-lived signals~\cite{stamp}, created, e.g., by rotational instabilities in nascent neutron stars~\cite{piro:07,piro:11,pirothrane12,corsi} or by the fragmentation of an accretion disk~\cite{kiuchi,vanputten:01,vanputten:08}, are potentially detectable by second-generation detectors~\cite{stochtrack,stochsky}.

Previous applications of seedless clustering have been employed to look for clusters of excess coherence using a coherent statistic~\cite{stochsky,verylong}:
\begin{equation}\label{eq:coherent}
  \mathfrak{p}(t;f) = \frac{2\sqrt{2}}{\cal N}
  \tilde{s}_I^*(t;f) \tilde{s}_J(t;f) \Big/ \sqrt{P'_I(t;f) P'_J(t;f)} .
\end{equation}
Here, $\tilde{s}_I(t;f)$ is the Fourier transform of the strain time series in detector $I$ for a data segment centered on time $t$ with frequencies $f$.
$P'_I(t;f)$ is the auto-power spectrum calculated using (typically nine) neighboring segments while ${\cal N}$ is a Fourier normalization constant.
By integrating $\mathfrak{p}(t;f)$ along suitable curves in frequency-time space (and applying a phase factor to ``point'' in different sky directions), one can define a signal-to-noise ratio for the cluster~\footnote{We correct the normalization factor given in Eq.~5 of~\cite{stochsky}: the coefficient is $1/N^{1/2}$ as opposed to $1/N$.}
\begin{equation}\label{eq:german_sum}
  \text{SNR}_\text{tot} \equiv
  \frac{1}{N^{1/2}}
  \sum_{\left\{t;f\right\}\in\Gamma}
  \text{Re}\left[
    e^{2\pi i f \Delta\tau} \mathfrak{p}(t;f)
    \right] .
\end{equation}
Here, $\Gamma$ is a template describing the set of spectrogram pixels in the parametrized (B\'ezier) curve, $N$ is the number of pixels in the curve, and $\Delta\tau$ is the direction-dependent time delay between two detectors.

By calculating $\text{SNR}_\text{tot}$ for many different templates $\Gamma$ and for many different time delays $\Delta\tau$, it is possible to find very weak signals buried in noise~\cite{stochsky}.
The loudest event in each $\mathfrak{p}(t;f)$ spectrogram is deemed to have a signal-to-noise ratio:
\begin{equation}
  \text{SNR}_\text{tot}^\text{max} = \max_\Gamma 
  \left[ \text{SNR}_\text{tot} \right]
\end{equation}
$\text{SNR}_\text{tot}^\text{max}$ is a coherent detection statistic.
The observed value of $\text{SNR}_\text{tot}^\text{max}$ is compared to a distribution of $\text{SNR}_\text{tot}^\text{max}$ generated using time-shifts in order to assign a false-alarm probability.

To reap the benefits of seedless clustering, it is desirable to employ a large number of templates: ${\cal O}(10^7)$ for a single $\approx$$\unit[300]{s}$ spectrogram
The calculation can be sped up by parallelization, but it becomes prohibitive to repeat $>10^6$ times on time-shifted data, the number of realizations required to carry out $5\sigma$ background estimation.
In order to eliminate this computational bottleneck, we introduce a single-detector, incoherent statistic
\begin{equation}\label{eq:lonetrack}
  \mathfrak{l}_I(t;f) = \frac{2\sqrt{2}}{\cal N}
  \left|\tilde{s}_I(t;f)\right|^2 \Big/ P'_I(t;f) .
\end{equation}
Note that $\mathfrak{l}_I(t;f)$ is equivalent to $\mathfrak{p}(t;f)$ for the case where the two detector indices are equal $I=J$.
It is the ratio of the auto-power in detector $I$ at time $t$ to the auto-power in detector $I$ at the times neighboring $t$.

While the seedless clustering algorithms described above were developed in the context of a coherent search using $\mathfrak{p}(t;f)$, it is straightforward to apply them to the new incoherent statistic $\mathfrak{l}_I(t;f)$ by simply defining a new single-detector signal-to-noise ratio:
\begin{equation}
  \text{SNR}_\text{tot}^{(I)} \equiv
  \frac{1}{N^{1/2}}
  \sum_{\left\{t;f\right\}\in\Gamma}
  \mathfrak{l}_I(t;f) .
\end{equation}
The phase factor from Eq.~\ref{eq:german_sum} vanishes because the time delay between two (now identical) detectors is zero.
As before, we identify the most significant cluster in each spectrogram.
It is necessary to do this separately for each detector: $\text{SNR}_\text{tot}^{\text{max } (I)} = \max_\Gamma \left[\text{SNR}_\text{tot}^{(I)}\right]$.

While $\mathfrak{l}_I(t;f)$ and $\mathfrak{p}_I(t;f)$ are similarly defined, it is important to note differences in their statistical behavior.
If no signal is present, the distribution of $\text{Re}\left[\mathfrak{p}(t;f)\right]$ is peaked symmetrically about zero.
However, the distribution of $\mathfrak{l}_I(t;f)$ is positive definite and asymmetric with a peak near one.
Thus, $\text{SNR}_\text{tot}^\text{max}$ [calculated from $\mathfrak{p}_I(t;f)$] and $\text{SNR}_\text{tot}^{\text{max }(I)}$ [calculated from $\mathfrak{l}_I(t;f)$] have very different distributions.
While $\text{SNR}_\text{tot}^\text{max}=10$ corresponds to a highly significant event, $\text{SNR}_\text{tot}^{\text{max }(I)}=29$ is typical for noise.

If we stopped here, and used $\text{SNR}_\text{tot}^{\text{max} (I)}$ as a detection statistic, the method would rely on an incoherent statistic.
However, we can do much better by using clusters identified with $\text{SNR}_\text{tot}^{\text{max} (I)}$ to calculate a coherent statistic.
The loudest cluster as ranked by the single-detector statistic is denoted $\Gamma_I$.
The coherent signal-to-noise ratio for this cluster is:
\begin{equation}
  \Lambda^{(I)}
  \equiv
  \frac{1}{N^{1/2}}
  \max_{\Delta\tau}
  \sum_{\left\{t;f\right\}\in\Gamma_I}
  \text{Re}\left[
    e^{2\pi i f \Delta\tau} \mathfrak{p}(t;f)
    \right] .
\end{equation}
The $\max_{\Delta\tau}$ term indicates that the sum is carried out for $400$ evenly spaced values of $\tau$, sufficient to match the diffraction-limited resolution.
{\em $\Lambda^{(I)}$ represents the coherent signal-to-noise ratio in detectors $I$ and $J$ for the loudest cluster identified using only the auto-power from detector $I$.}
Last, we define the detection statistic as the maximum value of $\Lambda^{(I)}$ among the two detectors: $\Lambda = \max_I \Lambda^{(I)}$ .

To illustrate why this hierarchical design is useful, it is helpful to describe the procedure of background estimation as a series of numbered steps.
\begin{enumerate}
\item We break the coincident data into conveniently long spectrograms to be analyzed for clusters. In this Letter, we use $50\%$-overlapping, $\approx$$\unit[300]{s}$ spectrograms.
\item For each spectrogram, we identify the loudest cluster $\Gamma_I$ in each detector using the single-detector, incoherent statistic $\text{SNR}_\text{tot}^{\text{max } (I)}$.
\item If $\text{SNR}_\text{tot}^{\text{max } (I)}$ is less than some pre-determined threshold $\text{SNR}_\text{th}^{(I)}$, proceed no further. The cluster is not promising enough to spend time calculating the coherent statistic.
\item For each spectrogram---if there is a cluster passing the cut $\text{SNR}_\text{tot}^{\text{max } (I)}<\text{SNR}_\text{th}^{(I)}$---we calculate the coherent detection statistic $\Lambda$.
\item Take the spectrogram data produced in step~1 and time-shift the clusters in one detector to create a new noise realization. Repeat steps 2--4 to generate a background distribution for $\Lambda$.
\end{enumerate}
Since steps~2--3 use a single-detector statistic, we obtain the same list of single-detector clusters every time---it does not matter if the data streams are shifted with respect to one another.
This means that we can carry out steps~2--3 just once and reuse the results for subsequent time-slides.
This is important because step~2 (cluster identification) is the computationally expensive step.
We still need to recalculate the coherent statistic $\Lambda$ for each time-slide, but this is a cheap calculation since we have reduced the number of templates from ${\cal O}(10^7)$ per spectrogram to one (and only a fraction of spectrograms will contain a cluster passing the step~3 cut).
The zero-lag data are analyzed identically with the same hierarchical process, ensuring that the background estimation can be used to identify detections.

{\em Mock data analysis.}
We carry out two mock data analyses.
For the first analysis, we analyze Monte Carlo noise.
The simulated noise is Gaussian, stationary in time, and colored according to the design sensitivity of Advanced LIGO~\cite{aligo}.
For the second analysis, we analyze time-shifted data from Initial LIGO, which has been recolored to resemble Advanced LIGO noise at design sensitivity.
The recolored noise preserves non-Gaussian noise artifacts that are present in real data, but not in Monte Carlo simulations.
% 29 September, 2015
Of course, the character of the non-Gaussian noise present in Advanced LIGO noise is likely to be different from that of Initial LIGO as the detectors are significantly different.
The recoloured noise results should therefore be taken as an plausible approximation of Advanced LIGO noise.
%%%%%%%%%%%%%%%%%%%%%
In both cases, we use a two-detector network consisting of the LIGO Hanford and Livingston observatories.

% (10000*288)/(3600*24)/2 = 16.7
%EHT: updated on 15 September, 2015 for one-sided spectrum
%We analyze $\unit[16.7]{days}$ of data, corresponding to $10^4$, $\unit[288]{s}$-long, $50\%$-overlapping spectrograms.
In order to estimate the background for one week of data, we analyze $\unit[16.7]{days}$ of data, corresponding to $10^4$, $\unit[288]{s}$-long, $50\%$-overlapping spectrograms.
The spectrograms span a frequency range of $100$--$\unit[1800]{Hz}$ and are composed of $\unit[1]{Hz}\times\unit[1]{s}$ pixels, overlapping $50\%$ in time.
This choice of spectrographic resolution is suitable for many long-lived transient waveforms with slowly varying frequency~\cite{pirothrane12,stochtrack,stochsky,lgrb}.

In order to impose the constraints of a realistic search, we assume that the seedless clustering parameters are tuned just once.
We employ cubic B\'ezier curves with a minimum duration of $\unit[40]{s}$.
For each spectrogram, we employ $10^7$ templates to find the loudest auto-power cluster for each detector $\Gamma_I$.
The difference in frequency between the first and third B\'ezier points determine the extent to which the signal may vary in time.
We limit this variation so that the frequency of the third control point is within $\pm50\%$ of the frequency of the first control point.

We consider three waveforms, which we have previously used to benchmark past studies using seedless clustering~\cite{stochtrack,stochsky}.
In particular, we employ two down-chirping accretion disk instability (ADI) signals from~\cite{lucia} and one up-chirping fallback accretion (FA) signal from~\cite{pirothrane12}.
Following~\cite{stochtrack,stochsky}, the accretion disk instability signals are scaled so that the isotropic equivalent energy is $0.1 M_\odot$.
The spectrographic properties of the signals are described in Tab.~\ref{tab:waveforms}.
We follow the naming conventions adopted in~\cite{stochtrack,stochsky}.

\begin{table}
  \begin{tabular}{|c|c|c|c|}
    \hline
    waveform & duration (s) & $f_\text{min}$--$f_\text{max}$ (Hz) & median strain \\\hline
    ADI 1 & $39$ & $130$--$170$ & $4 \times 10^{-24}$ \\\hline
    ADI 2 & $230$ & $110$--$260$ & $2 \times 10^{-23}$ \\\hline
    FA 2 & $200$ & $790$--$1080$ & $2 \times 10^{-22}$ \\\hline
  \end{tabular}
  \caption{
    Summary of the waveforms used in this study; see~\cite{stochtrack,stochsky} for additional details.
    Accretion disk instability (ADI) waveforms are down-chirping while fallback accretion (FA) waveforms are up-chirping.
    The ADI waveforms have been normalized so that the isotropic equivalent energy is $0.1M_\odot$.
    The median strain is for face-on systems at $\unit[100]{Mpc}$.
  }
  \label{tab:waveforms}
\end{table}

We do not include the results for a fourth waveform FA~1 from~\cite{stochtrack,stochsky}, which persists for $\unit[20]{s}$ and evolves from $790$--$\unit[1080]{Hz}$.
The implementation described above performed poorly detecting this waveform because it is relatively short and it evolves rapidly in frequency.
It is possible that we could effectively detect signals of this type using a different tuning from the one we describe above, but we seek to apply a uniform set of parameters, which are optimized for longer, less-quickly-evolving signals.

\begin{figure*}[hbtp!]
    % created with detect_coincidence_ltrack('adi-norm-d')
    % last updated on 13/05/15 at 10:10 PM
    % last updated on 06/06/15 at 12:11 PM
    % updated on 15/09/2015 using
    %   ~ethrane/lonetrack/F1000/mcoughlin/detect_coincidence_htrack_eric.m
    %   gsiscp $pcdev1:lonetrack/F1000/mcoughlin/lsnr1_htrack.eps lsnr1_rc.eps
    \subfigure[]{\psfig{file=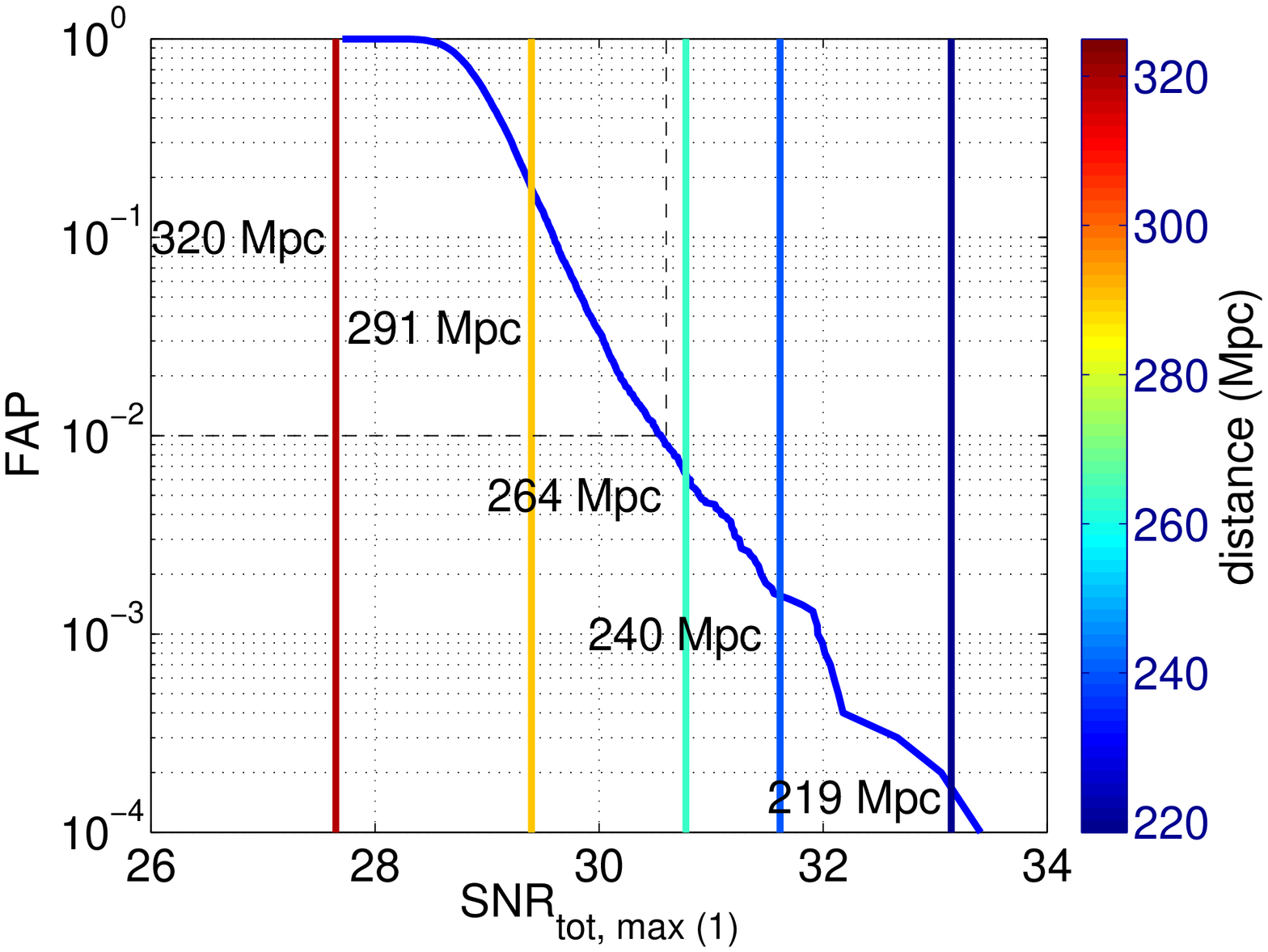, width=3in}}
    % created with detect_coincidence_htrack_bknd_v2('adi-norm-d')
    % last updated on 13/05/15 at 10:14 PM
    \subfigure[]{\psfig{file=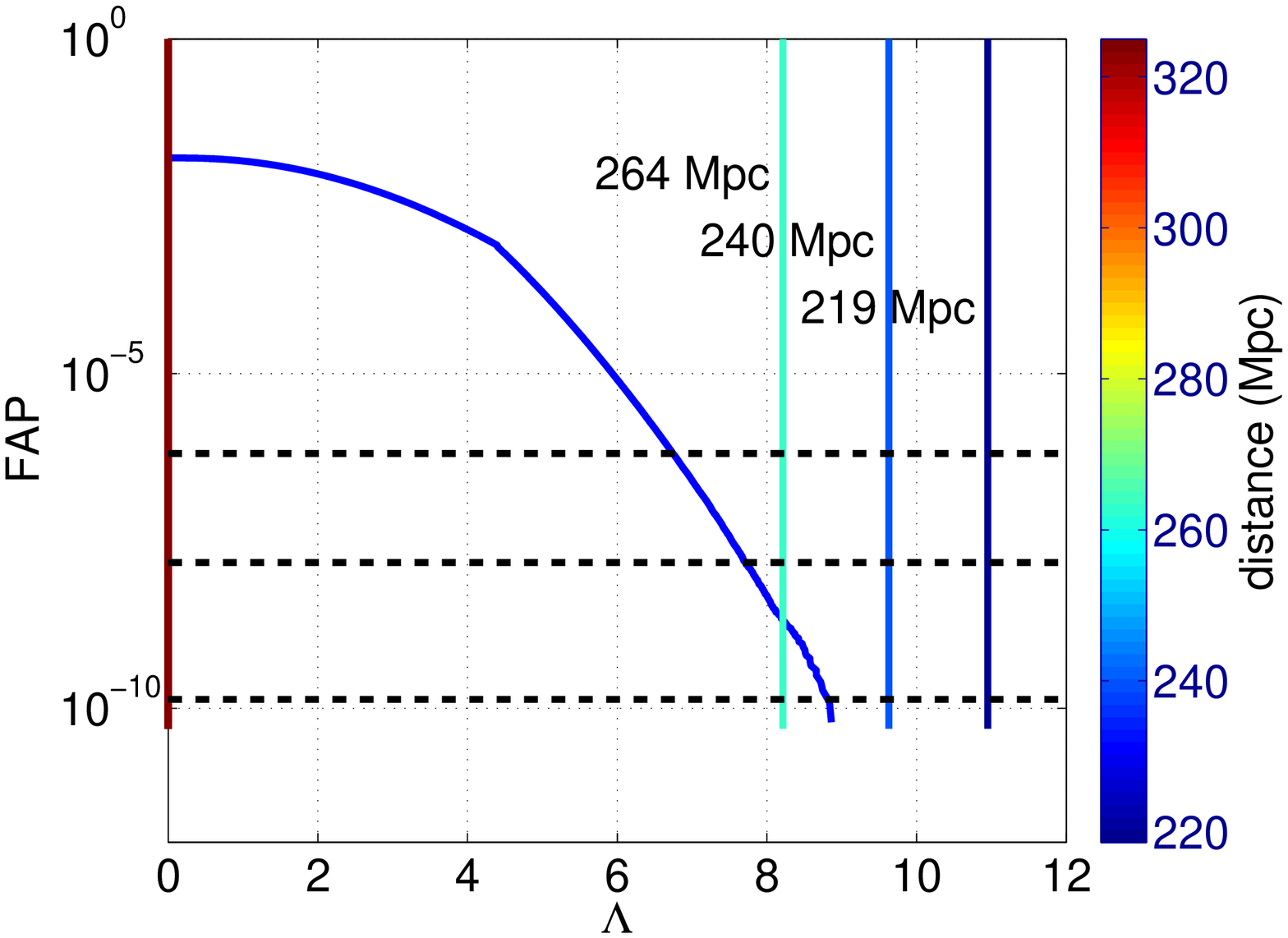, width=3in}}
    \caption{
      Left (a): FAP versus the single-detector statistic $\text{SNR}_\text{tot}^{\text{max }(I)}$.
      The blue curve indicates the distribution of recolored noise for Advanced LIGO at design sensitivity.
      The vertical lines show the median values for data including an accretion disk instability signal (ADI~2; face-on, optimal orientation) with different colors corresponding to different distances.
      The dashed black lines show which data are removed via the single-detector cut.
      Right (b): FAP versus the coherent statistic $\Lambda$.
      The blue curve indicates the distribution of recolored noise for Advanced LIGO (Hanford and Livingston) at design sensitivity.
      The vertical lines show the median values for data including an accretion disk instability signal (ADI~2; face-on, optimal orientation) with different colors corresponding to different distances.      
%EHT: updated on 15 September, 2015 for one-sided spectrum
%      The dashed black lines indicate the threshold required for $3\sigma$, $4\sigma$, and $5\sigma$ detection in two weeks of data.
      The dashed black lines indicate the threshold required for $3\sigma$, $4\sigma$, and $5\sigma$ detection in one week of data.
    }
    \label{fig:bknd}
\end{figure*}

Above, we listed five steps necessary for carrying out the hierarchical search.
The blue curve in Fig.~\ref{fig:bknd}a shows the FAP associated with the single-detector statistic $\text{SNR}_\text{tot}^{\text{max }(I)}$ obtained after steps 1--2 with recolored noise.
$\text{FAP}$ is defined for a single $\unit[288]{s}$ spectrogram so $\text{FAP}=10^{-2}$ corresponds to a false alarm rate of $\text{FAR}=\unit[(1/8)]{hr^{-1}}$.
The vertical lines in Fig.~\ref{fig:bknd}a show the median values of the single-detector statistic $\text{SNR}_\text{tot}^{\text{max }(I)}$ for data that includes an injected ADI~2 signal injected face-on at an optimal sky location.
The different colors indicate different source distances.

For step 3, it is necessary to define a threshold $\text{SNR}_\text{th}^{(I)}$ for the single-detector statistic.
This threshold is applied in order to control the computational costs.
By throwing out below-threshold candidates, we reduce the number of surviving clusters to include in subsequent steps.
The choice of $\text{SNR}_\text{th}^{(I)}$ is a balancing act.
Increasing the threshold reduces the computational burden, but it can also reduce sensitivity by inadvertently discarding true signals.
For the Monte Carlo analysis described here, we find that  $\text{SNR}_\text{th}^{(I)}=29.4$ eliminates $99\%$ of background triggers (thereby reducing the computational burden of steps 4--5 by a hundred fold) without severely harming sensitivity.
The step~3 cut is indicated by dashed black line on Fig.~\ref{fig:bknd}a.
Note that (face-on, optimally oriented) signals with distance $\lesssim$$\unit[260]{Mpc}$ are likely to survive the cut.
For recolored noise, the threshold is selected to be  $\text{SNR}_\text{th}^{(I)}=30.6$.

Having eliminated $99\%$ of the clusters, we proceed to calculate $\Lambda$.
In Fig.~\ref{fig:bknd}b, we plot FAP as a function of $\Lambda$ for recolored noise.
The different colored vertical lines indicate the median value of $\Lambda$ for injected signals (face-on, optimal orientation).
If an injected signal does not pass the step~3 auto-power cut in at least one detector, it is assigned a value of $\Lambda=0$.
The blue curve starts at $\text{FAP}\approx10^{-2}$ because $99\%$ of the clusters are removed by the auto-power cut.
%EHT: updated on 15 September, 2015 for one-sided spectrum
%The horizontal dashed lines indicate the required significance for $3\sigma$, $4\sigma$, and $5\sigma$ detections in two weeks of data.
The horizontal dashed lines indicate the required significance for $3\sigma$, $4\sigma$, and $5\sigma$ detections in one week of data.
% 2*14*24*3600 / 288 = 4200
% 1/(1.7e6*8400) = 7e-11
Recall that $\text{FAP}$ is measured in reference to a $\unit[288]{s}$ spectrogram.
%EHT: updated on 15 September, 2015 for one-sided spectrum
%Two weeks corresponds to $8400$ ($50\%$-overlapping) spectrograms and so $5\sigma$ corresponds to $\text{FAP}=7\times10^{-11}$.
One week corresponds to $4800$ ($50\%$-overlapping) spectrograms and so $5\sigma$ corresponds to $\text{FAP}=7\times10^{-11}$.

From Fig.~\ref{fig:bknd}b, it is apparent that the $\lesssim$$\unit[260]{Mpc}$ signals that we avoided cutting in step~3 (Fig.~\ref{fig:bknd}a) are detectable with the coherent $\Lambda$ statistic at the $>$$4\sigma$ level.
% ratio of FAPs = 6e-3 / 1e-9 
%Comparing Figs.~\ref{fig:bknd}a and~b, we can answer an interesting question: why not simply use the single-detector detection statistic?
From Figs.~\ref{fig:bknd}a and~b, we see why the hierarchical approach is superior to the single-detector statistic.
The coherent statistic yields a $\text{FAP}$ which is many orders of magnitude smaller.
Next, we consider how the sensitivity is affected by the use of an incoherent statistic as an intermediate step.
% ratio at 5 sigma: [10.8 8.6 8.6] ./ [7.9 7.6 6.4] = halftrack/stochsky
%    = 1.3671 1.2368    1.3437
% waveform: [ADI1 ADI2 PT2]
% 5 sigma distance: [150 139 18]
We compare $\Lambda$ to $\text{SNR}_\text{tot}^\text{max}$ for mock signals that are loud enough to detect with the hierarchical scheme at $5\sigma$ and find that they are comparable to within $37\%$.
Somewhat surprisingly, the hierarchical scheme is actually slightly more sensitive in this regime because it searches the sky in a systematic way whereas the $\text{SNR}_\text{tot}^\text{max}$ statistic pairs each track with a random sky position guess~\cite{stochsky}.

The initial clustering procedure (corresponding to steps 1--3) is carried out on Kepler GK104s Graphics Processor Units (GPUs) at a cost of $1.2$ continuously running GPUs.
% $\unit[348]{GPU\,s}/\unit[288]{s}$, implying a requirement of $1.2$ continuously running GPUs.
% http://ark.intel.com/products/64595/Intel-Xeon-Processor-E5-2670-20M-Cache-2_60-GHz-8_00-GTs-Intel-QPI
% 
We find that $18$ Intel Xeon E5-2670 Central Processing Unit (CPU) cores are required to match the performance of one GPU.
% -bash-4.1$ /home/ethrane/lonetrack/F1000/bin/run_halftrack_bknd 1
% Elapsed time is 843.845477 seconds.
%EHT: updated on 15 September, 2015 for one-sided spectrum
%Steps 4--5, in which we time-shift the data in order to calculate the coherent statistic $\Lambda$, are performed using single CPU cores at a cost of $2.9$ continuously running cores in order to achieve the background estimation necessary to identify $5\sigma$ detection candidates.
Steps 4--5, in which we time-shift the data in order to calculate the coherent statistic $\Lambda$, are performed using single CPU cores at a cost of $5.8$ continuously running cores in order to achieve the background estimation necessary to identify $5\sigma$ detection candidates.
The calculation in steps 4--5 does not, at present, benefit dramatically from GPU acceleration.
%EHT: updated on 15 September, 2015 for one-sided spectrum
%To put this in perspective, a fully coherent search, calculating $\text{SNR}_\text{tot}$ for each time-slide with seedless clustering would require $1.1\times10^6$ continuously running GPUs.
To put this in perspective, a fully coherent search, calculating $\text{SNR}_\text{tot}$ for each time-slide with seedless clustering would require $2.2\times10^6$ continuously running GPUs.

In Tab.~\ref{tab:results}, we summarize the results of a sensitivity study in which we estimate $5\sigma$ detection distance: the distance at which we can detect a waveform with a $\text{FAP}$ corresponding to $5\sigma$ with a false dismissal probability (FDP) of $50\%$.
We consider the cases of face-on, optimally oriented sources (useful for comparison with previous work~\cite{stochtrack,stochsky}) and also randomly oriented sources with random sky locations.
The first column is the waveform, the second is the type of noise (MC = Monte Carlo or RC = recolored noise), and the second is the $5\sigma$ detection distance.
The RC detection distances are sometimes smaller than the MC distances because non-stationary noise artifacts present in real data tend to decrease the sensitivity compared to idealized Gaussian noise.
%Tab.~\ref{tab:mc} shows the results for Monte Carlo while Tab.~\ref{tab:rn} is for recolored noise.

\begin{table}[h]
  \begin{tabular}{|l|c|c|}
    \hline
    Waveform                       & Noise & $5\sigma$ distance (Mpc) \\ \hline
    \multirow{2}{*}{ADI 1 optimal} & MC    & 250                      \\ \cline{2-3} 
                                   & RC    & 250                      \\ \hline
    \multirow{2}{*}{ADI 1 random}  & MC    & 150                      \\ \cline{2-3} 
                                   & RC    & 127                      \\ \hline
    \multirow{2}{*}{ADI 2 optimal} & MC    & 232                      \\ \cline{2-3} 
                                   & RC    & 232                      \\ \hline
    \multirow{2}{*}{ADI 2 random}  & MC    & 139                      \\ \cline{2-3} 
                                   & RC    & 130                      \\ \hline
    \multirow{2}{*}{PT 2 optimal}  & MC    & 30                       \\ \cline{2-3} 
                                   & RC    & 25                       \\ \hline
    \multirow{2}{*}{PT 2 random}   & MC    & 18                       \\ \cline{2-3} 
                                   & RC    & 14                       \\ \hline
  \end{tabular}
  \caption{
    Five sigma detection distances ($\text{FDP}=50\%$) for the three different test waveforms summarized in Table~\ref{tab:waveforms} in Gaussian Monte Carlo noise (MC) and Initial LIGO noise, recolored to match Advanced LIGO at design sensitivity (RC).
    We assume Hanford and Livingston detectors operating at design sensitivity.
    ``Optimal'' means that the source is face-on and optimally oriented to maximize detectability.
    ``Random'' means that the orientation and sky location of the source are chosen from random distributions.
  }
  \label{tab:results}
\end{table}

%EHT: updated on 15 September, 2015 for one-sided spectrum
%In order to put Tab.~\ref{tab:results} in context, we compare the face-on, optimal sky location, $5\sigma$ detection distances presented here---estimated for two weeks of data---to the $3.3\sigma$ detection distances from~\cite{stochsky}---estimated for a single $\unit[288]{s}$ spectrogram with a smaller $\unit[150]{Hz}$-wide band.
In order to put Tab.~\ref{tab:results} in context, we compare the face-on, optimal sky location, $5\sigma$ detection distances presented here---estimated for one week of data---to the $3.3\sigma$ detection distances from~\cite{stochsky}---estimated for a single $\unit[288]{s}$ spectrogram with a smaller $\unit[150]{Hz}$-wide band.
% 1/5.7e-7 = 1.8e06
% 8400 * (1.8e6/371) * (1700/150) = 
Despite the fact that the results here include an additional $4.6\times10^8$ trial factor, the hierarchical $5\sigma$ detection distances for Monte Carlo are only $0.45-0.82$ times less than the $3\sigma$ detection distances from~\cite{stochsky} obtained using seedless clustering.
For recolored noise the ratios are $0.50-0.96$.
The event rate and loudness of long-lived transients are unknown, but the sensitivity we demonstrate here is sufficient to potentially detect long-lived signals with second-generation detectors~\cite{pirothrane12,vanputten:08}.

% 29 September, 2015
% shortened sentence
For illustrative purposes, we focus here on the detection of $>$5$\sigma$ events with minimal resources.
As a result, the aggressive auto-power cut eliminates some fraction of $<$5$\sigma$ signals that could be detected with a different tuning.
However, the algorithm can be tuned differently to balance cost with sensitivity for different $\text{FAP}$s.
For future work, it is worthwhile considering if this strategy can be usefully employed in other situations where the character of the noise is less well-behaved.
%%%%%%%%%%%%%%%%%%%%
% 29 September, 2015
Moreover, the general strategy we outline here---splitting a search into a computationally expensive ``incoherent'' step followed by a computationally cheap ``coherent'' step in order to facilitate rapid background estimation---may find use in the broader community.
For example, a similar scheme could prove useful in determining the significance of potentially faint electromagnetic signatures found (by $\geq2$ telescopes) in coincidence with gravitational-wave detections.
%%%%%%%%%%%%%%%%%%%%
We thank Anthony Piro for sharing the fallback accretion waveforms used in this analysis.
We thank Peter Shawhan and Jonah Kanner for helpful comments.

\bibliography{lonetrack}

\begin{thebibliography}{24}
\expandafter\ifx\csname natexlab\endcsname\relax\def\natexlab#1{#1}\fi
\expandafter\ifx\csname bibnamefont\endcsname\relax
  \def\bibnamefont#1{#1}\fi
\expandafter\ifx\csname bibfnamefont\endcsname\relax
  \def\bibfnamefont#1{#1}\fi
\expandafter\ifx\csname citenamefont\endcsname\relax
  \def\citenamefont#1{#1}\fi
\expandafter\ifx\csname url\endcsname\relax
  \def\url#1{\texttt{#1}}\fi
\expandafter\ifx\csname urlprefix\endcsname\relax\def\urlprefix{URL }\fi
\providecommand{\bibinfo}[2]{#2}
\providecommand{\eprint}[2][]{\url{#2}}

\bibitem[{\citenamefont{Abadie et~al.}(2012{\natexlab{a}})}]{s6lowmass}
\bibinfo{author}{\bibfnamefont{J.}~\bibnamefont{Abadie}} \bibnamefont{et~al.},
  \bibinfo{journal}{Phys. Rev. D} \textbf{\bibinfo{volume}{85}},
  \bibinfo{pages}{082002} (\bibinfo{year}{2012}{\natexlab{a}}).

\bibitem[{\citenamefont{Was et~al.}(2010)\citenamefont{Was, Bizouard, Brisson,
  Cavalier, Davier, Hello, Leroy, Robinet, and
  Vavoulidis}}]{0264-9381-27-1-015005}
\bibinfo{author}{\bibfnamefont{M.}~\bibnamefont{Was}},
  \bibinfo{author}{\bibfnamefont{M.-A.} \bibnamefont{Bizouard}},
  \bibinfo{author}{\bibfnamefont{V.}~\bibnamefont{Brisson}},
  \bibinfo{author}{\bibfnamefont{F.}~\bibnamefont{Cavalier}},
  \bibinfo{author}{\bibfnamefont{M.}~\bibnamefont{Davier}},
  \bibinfo{author}{\bibfnamefont{P.}~\bibnamefont{Hello}},
  \bibinfo{author}{\bibfnamefont{N.}~\bibnamefont{Leroy}},
  \bibinfo{author}{\bibfnamefont{F.}~\bibnamefont{Robinet}}, \bibnamefont{and}
  \bibinfo{author}{\bibfnamefont{M.}~\bibnamefont{Vavoulidis}},
  \bibinfo{journal}{Class. Quant. Grav.} \textbf{\bibinfo{volume}{27}},
  \bibinfo{pages}{015005} (\bibinfo{year}{2010}).

\bibitem[{\citenamefont{{Sutton P. et al.}}(2010)}]{X-Pipeline}
\bibinfo{author}{\bibnamefont{{Sutton P. et al.}}}, \bibinfo{journal}{New
  Journal of Physics} \textbf{\bibinfo{volume}{12}}, \bibinfo{pages}{053034}
  (\bibinfo{year}{2010}).

\bibitem[{\citenamefont{Klimenko et~al.}(2008)\citenamefont{Klimenko, Yakushin,
  Mercer, and Mitselmakher}}]{CoherentWaveBurst}
\bibinfo{author}{\bibfnamefont{S.}~\bibnamefont{Klimenko}},
  \bibinfo{author}{\bibfnamefont{I.}~\bibnamefont{Yakushin}},
  \bibinfo{author}{\bibfnamefont{A.}~\bibnamefont{Mercer}}, \bibnamefont{and}
  \bibinfo{author}{\bibfnamefont{S.}~\bibnamefont{Mitselmakher}},
  \bibinfo{journal}{Class. Quant. Grav.} \textbf{\bibinfo{volume}{25}},
  \bibinfo{pages}{114029} (\bibinfo{year}{2008}).

\bibitem[{\citenamefont{Thrane et~al.}(2011)\citenamefont{Thrane, Kandhasamy,
  Ott et~al.}}]{stamp}
\bibinfo{author}{\bibfnamefont{E.}~\bibnamefont{Thrane}},
  \bibinfo{author}{\bibfnamefont{S.}~\bibnamefont{Kandhasamy}},
  \bibinfo{author}{\bibfnamefont{C.~D.} \bibnamefont{Ott}},
  \bibnamefont{et~al.}, \bibinfo{journal}{Phys. Rev. D}
  \textbf{\bibinfo{volume}{83}}, \bibinfo{pages}{083004}
  (\bibinfo{year}{2011}).

\bibitem[{\citenamefont{Abadie et~al.}(2012{\natexlab{b}})}]{s6vsr23_allsky}
\bibinfo{author}{\bibfnamefont{J.}~\bibnamefont{Abadie}} \bibnamefont{et~al.},
  \bibinfo{journal}{Phys. Rev. D} \textbf{\bibinfo{volume}{85}},
  \bibinfo{pages}{122007} (\bibinfo{year}{2012}{\natexlab{b}}).

\bibitem[{\citenamefont{Anderson et~al.}(2001)\citenamefont{Anderson, Brady,
  Creighton, and Flanagan}}]{box}
\bibinfo{author}{\bibfnamefont{W.~G.} \bibnamefont{Anderson}},
  \bibinfo{author}{\bibfnamefont{P.~R.} \bibnamefont{Brady}},
  \bibinfo{author}{\bibfnamefont{J.~D.~E.} \bibnamefont{Creighton}},
  \bibnamefont{and} \bibinfo{author}{\bibfnamefont{{\'E}.~{\'E}.}
  \bibnamefont{Flanagan}}, \bibinfo{journal}{Phys. Rev. D}
  \textbf{\bibinfo{volume}{63}}, \bibinfo{pages}{042003}
  (\bibinfo{year}{2001}).

\bibitem[{\citenamefont{Abadie et~al.}(2010)}]{s5vsr1_allsky}
\bibinfo{author}{\bibfnamefont{J.}~\bibnamefont{Abadie}} \bibnamefont{et~al.},
  \bibinfo{journal}{Phys. Rev. D} \textbf{\bibinfo{volume}{81}},
  \bibinfo{pages}{102001} (\bibinfo{year}{2010}).

\bibitem[{\citenamefont{Thrane and Coughlin}(2013)}]{stochtrack}
\bibinfo{author}{\bibfnamefont{E.}~\bibnamefont{Thrane}} \bibnamefont{and}
  \bibinfo{author}{\bibfnamefont{M.}~\bibnamefont{Coughlin}},
  \bibinfo{journal}{Phys. Rev. D} \textbf{\bibinfo{volume}{88}},
  \bibinfo{pages}{083010} (\bibinfo{year}{2013}).

\bibitem[{\citenamefont{Thrane and Coughlin}(2014)}]{stochsky}
\bibinfo{author}{\bibfnamefont{E.}~\bibnamefont{Thrane}} \bibnamefont{and}
  \bibinfo{author}{\bibfnamefont{M.}~\bibnamefont{Coughlin}},
  \bibinfo{journal}{Phys. Rev. D} \textbf{\bibinfo{volume}{89}},
  \bibinfo{pages}{063012} (\bibinfo{year}{2014}).

\bibitem[{\citenamefont{Aasi et~al.}(2015)}]{aligo}
\bibinfo{author}{\bibfnamefont{J.}~\bibnamefont{Aasi}} \bibnamefont{et~al.},
  \bibinfo{journal}{Class. Quant. Grav.} \textbf{\bibinfo{volume}{32}},
  \bibinfo{pages}{074001} (\bibinfo{year}{2015}).

\bibitem[{\citenamefont{Thrane et~al.}(2015)\citenamefont{Thrane, Mandic, and
  Christensen}}]{verylong}
\bibinfo{author}{\bibfnamefont{E.}~\bibnamefont{Thrane}},
  \bibinfo{author}{\bibfnamefont{V.}~\bibnamefont{Mandic}}, \bibnamefont{and}
  \bibinfo{author}{\bibfnamefont{N.}~\bibnamefont{Christensen}},
  \bibinfo{journal}{Phys. Rev. D} \textbf{\bibinfo{volume}{91}},
  \bibinfo{pages}{104021} (\bibinfo{year}{2015}).

\bibitem[{\citenamefont{Coughlin et~al.}(2014)\citenamefont{Coughlin, Thrane,
  and Christensen}}]{stochtrack_cbc}
\bibinfo{author}{\bibfnamefont{M.}~\bibnamefont{Coughlin}},
  \bibinfo{author}{\bibfnamefont{E.}~\bibnamefont{Thrane}}, \bibnamefont{and}
  \bibinfo{author}{\bibfnamefont{N.}~\bibnamefont{Christensen}},
  \bibinfo{journal}{Phys. Rev. D} \textbf{\bibinfo{volume}{90}},
  \bibinfo{pages}{083005} (\bibinfo{year}{2014}).

\bibitem[{\citenamefont{Coughlin et~al.}(2015)\citenamefont{Coughlin, Meyers,
  Thrane, Luo, and Christensen}}]{stochtrack_ecbc}
\bibinfo{author}{\bibfnamefont{M.}~\bibnamefont{Coughlin}},
  \bibinfo{author}{\bibfnamefont{P.}~\bibnamefont{Meyers}},
  \bibinfo{author}{\bibfnamefont{E.}~\bibnamefont{Thrane}},
  \bibinfo{author}{\bibfnamefont{J.}~\bibnamefont{Luo}}, \bibnamefont{and}
  \bibinfo{author}{\bibfnamefont{N.}~\bibnamefont{Christensen}},
  \bibinfo{journal}{Phys. Rev. D} \textbf{\bibinfo{volume}{91}},
  \bibinfo{pages}{063004} (\bibinfo{year}{2015}).

\bibitem[{\citenamefont{Farin}(1996)}]{bezier}
\bibinfo{author}{\bibfnamefont{G.}~\bibnamefont{Farin}},
  \emph{\bibinfo{title}{Curves and Surfaces for CAGD, Fourth Edition: A
  Practical Guide}} (\bibinfo{publisher}{Academic Press},
  \bibinfo{year}{1996}).

\bibitem[{\citenamefont{Piro and Pfahl}(2007)}]{piro:07}
\bibinfo{author}{\bibfnamefont{A.~L.} \bibnamefont{Piro}} \bibnamefont{and}
  \bibinfo{author}{\bibfnamefont{E.}~\bibnamefont{Pfahl}},
  \bibinfo{journal}{Astrophys. J.} \textbf{\bibinfo{volume}{658}},
  \bibinfo{pages}{1173} (\bibinfo{year}{2007}).

\bibitem[{\citenamefont{Piro and Ott}(2011)}]{piro:11}
\bibinfo{author}{\bibfnamefont{A.~L.} \bibnamefont{Piro}} \bibnamefont{and}
  \bibinfo{author}{\bibfnamefont{C.~D.} \bibnamefont{Ott}},
  \bibinfo{journal}{Astrophys. J.} \textbf{\bibinfo{volume}{736}},
  \bibinfo{pages}{108} (\bibinfo{year}{2011}).

\bibitem[{\citenamefont{Piro and Thrane}(2012)}]{pirothrane12}
\bibinfo{author}{\bibfnamefont{A.~L.} \bibnamefont{Piro}} \bibnamefont{and}
  \bibinfo{author}{\bibfnamefont{E.}~\bibnamefont{Thrane}},
  \bibinfo{journal}{Astrophys. J.} \textbf{\bibinfo{volume}{761}},
  \bibinfo{pages}{63} (\bibinfo{year}{2012}).

\bibitem[{\citenamefont{Corsi and M\'esz\'aros}(2009)}]{corsi}
\bibinfo{author}{\bibfnamefont{A.}~\bibnamefont{Corsi}} \bibnamefont{and}
  \bibinfo{author}{\bibfnamefont{P.}~\bibnamefont{M\'esz\'aros}},
  \bibinfo{journal}{Astrophys. J.} \textbf{\bibinfo{volume}{702}},
  \bibinfo{pages}{1171} (\bibinfo{year}{2009}).

\bibitem[{\citenamefont{Kiuchi et~al.}(2011)\citenamefont{Kiuchi, Shibata,
  Montero, and Font}}]{kiuchi}
\bibinfo{author}{\bibfnamefont{K.}~\bibnamefont{Kiuchi}},
  \bibinfo{author}{\bibfnamefont{M.}~\bibnamefont{Shibata}},
  \bibinfo{author}{\bibfnamefont{P.~J.} \bibnamefont{Montero}},
  \bibnamefont{and} \bibinfo{author}{\bibfnamefont{J.~A.} \bibnamefont{Font}},
  \bibinfo{journal}{Phys. Rev. Lett.} \textbf{\bibinfo{volume}{106}},
  \bibinfo{pages}{251102} (\bibinfo{year}{2011}).

\bibitem[{\citenamefont{van Putten}(2001)}]{vanputten:01}
\bibinfo{author}{\bibfnamefont{M.~H. P.~M.} \bibnamefont{van Putten}},
  \bibinfo{journal}{Phys. Rev. Lett.} \textbf{\bibinfo{volume}{87}},
  \bibinfo{pages}{091101} (\bibinfo{year}{2001}).

\bibitem[{\citenamefont{van Putten}(2008)}]{vanputten:08}
\bibinfo{author}{\bibfnamefont{M.~H. P.~M.} \bibnamefont{van Putten}},
  \bibinfo{journal}{Astrophys. J. Lett.} \textbf{\bibinfo{volume}{684}},
  \bibinfo{pages}{L91} (\bibinfo{year}{2008}).

\bibitem[{\citenamefont{Aasi et~al.}(2013)}]{lgrb}
\bibinfo{author}{\bibfnamefont{J.}~\bibnamefont{Aasi}} \bibnamefont{et~al.},
  \bibinfo{journal}{Phys. Rev. D} \textbf{\bibinfo{volume}{88}},
  \bibinfo{pages}{122004} (\bibinfo{year}{2013}).

\bibitem[{\citenamefont{Santamar\'ia and Ott}(2011)}]{lucia}
\bibinfo{author}{\bibfnamefont{L.}~\bibnamefont{Santamar\'ia}}
  \bibnamefont{and} \bibinfo{author}{\bibfnamefont{C.~D.} \bibnamefont{Ott}},
  \emph{\bibinfo{title}{Gravitational wave emission from accretion disk
  instabilities - analytic models}} (\bibinfo{year}{2011}),
  \bibinfo{note}{\url{https://dcc.ligo.org/LIGO-T1100093-v2/public}}.

\end{thebibliography}

\end{document}